\documentclass[11pt,a4paper]{article}

\usepackage{jheppub}

\pdfoutput=1

\usepackage{amsmath,amssymb}
\usepackage{mathrsfs}
\usepackage{comment}
\usepackage{subcaption}

\allowdisplaybreaks

\title{Quantum fermion superradiance and vacuum ambiguities on charged black holes}

\author[a]{\'{A}lvaro \'{A}lvarez-Dom\'{\i}nguez}
\author[b]{and Elizabeth Winstanley}

\affiliation[a]{Departamento de Física Te\'{o}rica and IPARCOS, \\ 
Universidad Complutense de Madrid, 
Plaza de las Ciencias 1, 
28040 Madrid, Spain.}

\affiliation[b]{School of Mathematical and Physical Sciences,
	The University of Sheffield,\\
	Hicks Building,
	Hounsfield Road,
	Sheffield. S3 7RH United Kingdom.}

\emailAdd{alvalv04@ucm.es}

\emailAdd{E.Winstanley@sheffield.ac.uk}

\abstract{Unlike a classical charged bosonic field, a classical charged fermion field on a static charged black hole does not exhibit superradiant scattering. We demonstrate that the quantum analogue of this classical process is however present. We construct a vacuum state for the fermion field which has no incoming particles from past null infinity, but which contains, at future null infinity, a nonthermal flux of particles. This state describes both the discharge and energy loss of the black hole, and we analyze how the interpretation of this phenomenon depends on the ambiguities inherent in defining the quantum vacuum. }


\preprint{IPARCOS-UCM-24-055}
\arxivnumber{2411.00167 [hep-th]}

\begin{document}

\maketitle
\flushbottom

\section{Introduction}
\label{sec:intro}

Superradiance \cite{Brito:2015oca} is a classical phenomenon whereby a field wave is amplified during a scattering process.
In black hole physics, superradiance arises when low-frequency bosonic field waves are scattered on a rotating black hole \cite{Misner:1972kx,Press:1972zz,Teukolsky:1974yv,Chandrasekhar:1985kt}. 
For bosonic fields, superradiance is a consequence of the area theorem and first law of black hole mechanics \cite{Bekenstein:1973mi}, the former holding for matter fields satisfying the weak energy condition. 
Superradiance does not occur for classical fermion fields on rotating black hole backgrounds \cite{Unruh:1973bda,Chandrasekhar:1976ap,Iyer:1978du} because they do not satisfy the weak energy condition, and so the area law no longer holds \cite{Unruh:1973bda}.

Both bosonic and fermionic fields do however exhibit the quantum analogue of classical superradiance \cite{Starobinsky:1973aij,Unruh:1974bw,Dai:2023zcj,Dai:2023ewf}.
Particles are spontaneously emitted in low-frequency field modes, in precisely those frequencies which correspond to classically superradiant modes for bosonic fields. 
The radiation is nonthermal in nature and is in addition to the usual Hawking radiation \cite{Hawking:1975vcx} emitted by the black hole. 

Classical superradiance also occurs on static, nonrotating black holes when both the black hole and the scattered field have a nonzero charge (`charge superradiance')~\cite{Bekenstein:1973mi,Nakamura:1976nc,DiMenza:2014vpa,Benone:2015bst,DiMenza:2019zli}.
As with the classical superradiance of neutral fields on rotating black holes, charge superradiance only exists for bosonic and not fermionic fields~\cite{Maeda:1976tm}.
A natural question is whether there is a quantum analogue of this classical charge superradiance. 
For a  massless charged scalar field, this process was studied many years ago  \cite{Gibbons:1975kk} and revisited more recently~\cite{Balakumar:2020gli}. 
In~\cite{Balakumar:2020gli}, `in' and `out' vacuum states are constructed for the charged quantum scalar field on a Reissner-Nordstr\"om~(RN) black hole. 
The `in' state is devoid of particles at past null infinity, but contains an outgoing flux of particles at future null infinity. 
This flux is present only in those modes which exhibit classical charge superradiance. 

Our purpose in this paper is to investigate whether the quantum analogue of charge superradiance also occurs for a massless charged fermion field. We construct analogues of the `in' and `out' states defined for a charged scalar field in~\cite{Balakumar:2020gli}. These quantum vacua describe the discharge and energy loss of the charged black hole, leading to a dissipative phenomenon that is significantly more intense than in the scalar case.

However, we show that quantum superradiance is not exhibited by all fermionic quantum states that can be defined in RN black holes. In particular, we construct a candidate `Boulware' state---originally introduced for scalar fields on Schwarzschild black holes~\cite{Boulware:1975}---which exhibits no particle flux at either past or future null infinity. 
This marks a significant distinction from scalar fields on RN black holes, where, due to additional restrictions in the canonical quantization process, a direct analogue of the `Boulware' state does not exist~\cite{Balakumar:2022}.

We introduce the set-up in section~\ref{sec:classical} and we review the separable solutions of the charged Dirac equation on RN in section~\ref{sec:modes}. 
In section~\ref{sec:quantum} we discuss the ambiguities in the definition of quantum vacuum, and we construct the `in' and `out' states that describe the phenomenon of quantum superradiance. 
In section~\ref{sec:quantumsuperradiance} we examine this effect, calculating the number density of created particles as well as the black hole discharge and energy loss. 
Section~\ref{sec:conc} contains further discussion and our conclusions.

\section{Classical charged fermion field on a charged black hole}
\label{sec:classical}

We consider a massless charged fermion field propagating on a background charged Reissner-Nordstr\"om (RN) black hole.
The spacetime is described by the line element
\begin{equation}
\text{d}s^{2} = - f(r) \, \text{d}t^{2} + \left[ f(r) \right] ^{-1} \text{d}r^{2}+ r^{2} \text{d}\theta ^{2} + r^{2}\sin ^{2} \theta \, \text{d}\varphi ^{2} ,
\label{eq:RNmetric}
\end{equation}
where the metric function $f(r)$ is given by 
\begin{equation}
f(r) = 1 - \frac{2M}{r} + \frac{Q^{2}}{r^{2}} ,
\label{eq:fr}
\end{equation}
with $M$ being the mass and $Q$ the electric charge of the black hole.
We restrict attention to the situation where $M^{2}>Q^{2}$,
in which case
the metric function $f(r)$ has two zeros at $r=r_{\pm }$, where
\begin{equation}
r_{\pm } = M \pm {\sqrt {M^{2}-Q^{2}}}.
\label{eq:rpm}
\end{equation}
The larger root~$r_{+}$ is the location of the black hole event horizon 
and $r_{-}$ is the location of the inner horizon.
We will be interested only in the region of spacetime exterior to the event horizon. 

A massless fermionic field~$\Psi$ with charge $q$ propagating on the RN black hole satisfies the Dirac equation
\begin{equation}
\gamma ^{\mu }\left( \nabla _{\mu } + iqA_{\mu } \right) \Psi = 0.
\label{eq:Dirac}
\end{equation}
This equation involves the spinor covariant derivative~$\nabla_\mu$ and the electromagnetic gauge potential~$A_\mu$, where (with an appropriate choice of gauge)
\begin{equation}
A_\mu = \left(-\frac{Q}{r}, 0, 0, 0\right).
\label{eq:gaugepot}
\end{equation}
The suitable basis chosen for the~$\gamma^\mu$ matrices, as well as the complete expression for~$\nabla^\mu$ in terms of the spinor connection matrices, is provided in Appendix~\ref{sec:useful}. 
The resulting form for the Dirac equation \eqref{eq:Dirac} is then given by 
\begin{multline}
\left\{  \widetilde{\gamma}^0\frac{1}{\sqrt{f(r)}}\left(\partial_t-i\frac{qQ}{r}\right)
    +  \widetilde{\gamma}^3\sqrt{f(r)}\left[\partial_r+\frac{1}{4f(r)}\frac{\text{d}f}{\text{d}r} +\frac{1}{r}\right] 
    \right. \\ \left.
    +  \widetilde{\gamma}^1\frac{1}{r}\left[ \partial_\theta + \frac{\cot\theta}{2}\right]
    +  \widetilde{\gamma}^2\frac{1}{r\sin\theta}\partial_\phi \right\} \Psi = 0.
    \label{eq:DiracCoord}
\end{multline}

For any two solutions $\Psi_1$ and $\Psi_2$ of the Dirac equation~\eqref{eq:Dirac}, we can define the positive-definite inner product
\begin{equation}
\left( \Psi _{1}, \Psi _{2} \right) = \int _{\Sigma_{t}} {\overline {\Psi }}_{1}
\gamma ^{\mu }n_{\mu } \Psi _{2} \, \text{d}\Sigma,
\label{eq:innerproduct}
\end{equation}
where ${\overline{\Psi }}={\widetilde{\gamma }}^{0}\Psi ^{\dagger}$ (with $\Psi ^{\dagger}$ denoting the Hermitian conjugate of $\Psi $) and $n_{\mu }$ is the unit outwards-pointing normal to $\Sigma_{t}$. This inner product is independent of the constant $t$ hypersurface $\Sigma_t$ on which it is evaluated, thereby endowing the space of Dirac solutions with a Hilbert space structure. 
This is different from the case of a charged scalar field propagating on a RN background \cite{Balakumar:2022}. 
The space of Klein-Gordon solutions lacks a naturally defined inner product, instead having an antisymmetric symplectic form that is not positive definite \cite{Balakumar:2022}. 
In other words,  charged scalar field modes can have positive or negative norm, while charged fermionic modes can only have positive norm. 
This difference will influence canonical quantization, providing greater flexibility in the fermionic case compared to the scalar case.

\section{Solutions of the Dirac equation and classical superradiance}
\label{sec:modes}

We search for a separable orthonormal basis of solutions~$\{\psi_\Lambda\}$ of the Dirac equation~\eqref{eq:DiracCoord} with respect to the inner product~\eqref{eq:innerproduct}. 
Massless fermion solutions can be classified according to the equation~\cite{Unruh:1973bda,Iyer:1982ah} 
\begin{equation}
\left( 1 - L \gamma ^{5} \right) \Psi  = 0,
\label{eq:lefthanded}
\end{equation}
where~$\gamma^5$ is the chirality matrix (see Appendix~\ref{sec:useful}). 
Spinors with $L=1$ are called `left-handed' and those with $L=-1$, `right-handed'. 
For a given chirality $L$, we propose the ansatz \cite{Unruh:1973bda,Unruh:1974bw,Vilenkin:1978is}
\begin{equation}
    \psi_\Lambda (t,r,\theta,\varphi) = \frac{(\sqrt{8\pi^2})^{-1}}{\mathcal{F}(r,\theta)}e^{-i\omega t}e^{im\varphi} 
    \left(
\begin{array}{c}
\eta_\Lambda(r,\theta) \\
L\eta_\Lambda(r,\theta)
\end{array}
\right),
\label{eq:ansatz}
\end{equation}
where 
\begin{equation}
\mathcal{F}(r,\theta)=r[f(r)\sin^2\theta]^{1/4},
\end{equation}
for each mode labelled by a set of quantum indices \mbox{$\Lambda=\{\omega,l,m\}$}. 
Each component of the two-spinors~$\eta_\Lambda(r,\theta)$ is separable into radial and angular functions:
\begin{equation}
    \eta_\Lambda(r,\theta) = 
    \left(
\begin{array}{c}
R_{1,\Lambda}(r) S_{1,\Lambda}(\theta) \\
R_{2,\Lambda}(r) S_{2,\Lambda}(\theta)
\end{array}
\right).
\end{equation}
Introducing this ansatz into the Dirac equation~\eqref{eq:DiracCoord}, we find two linearly independent equations for the angular functions: 
\begin{align}
\label{eq:S12}
    \left[ \frac{\text{d}}{\text{d}\theta} - \frac{m}{\sin\theta} \right] S_{1,\Lambda}(\theta) & = \left(l+\frac{1}{2}\right) S_{2,\Lambda}(\theta), \nonumber \\
    \left[ \frac{\text{d}}{\text{d}\theta} + \frac{m}{\sin\theta} \right] S_{2,\Lambda}(\theta) & = -\left(l+\frac{1}{2}\right) S_{1,\Lambda}(\theta),
\end{align}
as well as for the radial functions:
\begin{align}
\label{eq:R12}
    r\sqrt{f(r)}\left[ \frac{\text{d}}{\text{d}r} -\frac{iL}{f(r)}\left( \omega+\frac{qQ}{r} \right)\right] R_{1,\Lambda}(r) & = \left(l+\frac{1}{2}\right) R_{2,\Lambda}(r), \nonumber \\
    r\sqrt{f(r)}\left[ \frac{\text{d}}{\text{d}r} + \frac{iL}{f(r)}\left( \omega+\frac{qQ}{r} \right)\right] R_{2,\Lambda}(r) & = \left(l+\frac{1}{2}\right) R_{1,\Lambda}(r).
\end{align}
These functions have a discrete spectrum, with \mbox{$l=\frac{1}{2}, \frac{3}{2}, ...$}; and \mbox{$m= -l, -l+1, ..., l-1, l$}. 
We take the angular functions to be real.
They are related to the well-known spin-weighted spherical harmonics~\cite{Newman:1966ub,Goldberg:1966uu}, and
we give some of their properties in Appendix~\ref{sec:propertiesRS}.

Writing the radial equations~\eqref{eq:R12} in terms of the tortoise coordinate~$r_*$, defined by
\begin{equation}
    \frac{\text{d}r_*}{\text{d}r} = \frac{1}{f(r)},
\end{equation}
one can verify that the radial functions behave as plane waves asymptotically far from the black hole ($r_*\rightarrow \infty$):
\begin{equation}
    R_{1,\Lambda}(r_*) \propto e^{iL\omega r_*}, \qquad R_{2,\Lambda}(r_*) \propto e^{-iL\omega r_*}.
    \label{eq:R-infty}
\end{equation}
This is as expected, since the RN spacetime is asymptotically flat. At the event horizon~$r_+$ ($r_* \rightarrow -\infty$), these functions also behave as plane waves,
\begin{equation}
    R_{1,\Lambda}(r_*) \propto e^{iL\widetilde{\omega} r_*}, \qquad R_{2,\Lambda}(r_*) \propto e^{-iL\widetilde{\omega} r_*},
    \label{eq:R+infty}
\end{equation}
but now with a shifted frequency~
\begin{equation}
    \widetilde{\omega}=\omega+\frac{qQ}{r_+}.
    \label{eq:omegatilde}
\end{equation}
For positive chirality $L=1$, the plane waves of $R_{1,\Lambda}(r_*)$ are outgoing at both future null infinity~$\mathscr{I}^+$ and the past event horizon~$\mathscr{H}^-$, while those of $R_{2,\Lambda}(r_*)$ are ingoing at both past null infinity~$\mathscr{I}^-$ and the future event horizon~$\mathscr{H}^+$. 
For negative chirality $L=-1$, the roles of $R_{1,\Lambda}(r_*)$ and $R_{2,\Lambda}(r_*)$ are reversed. 
Therefore, in what follows, we will restrict our attention to the case of positive chirality, $L=1$.

We now define two well-known orthonormal bases of solutions to the Dirac equation. 
Elements of the bases will be of the form given by~\eqref{eq:ansatz}, with certain choices of the radial functions. 

The so-called `in-up' basis is determined by imposing initial conditions for the radial functions on the past hypersurface~$\mathscr{H}^- \cup \mathscr{I}^-$~\footnote{Strictly speaking, we are choosing a Cauchy surface close to~$\mathscr{H}^- \cup \mathscr{I}^-$.}. 
The `in' modes represent unit flux of incoming plane waves from~$\mathscr{I}^-$, with no contribution coming from~$\mathscr{H}^-$. 
These ingoing plane waves are partly transmitted to the future horizon~$\mathscr{H}^+$ and, since this is a scattering problem, partly reflected back to~$\mathscr{I}^+$. 
According to~(\ref{eq:R-infty}, \ref{eq:R+infty}), this translates into the asymptotic behaviour of the radial functions given by
\begin{align}
    (R_{1,\Lambda}^{\text{in}},R_{2,\Lambda}^{\text{in}}) \sim 
    \begin{cases}
    (0,t_\Lambda^{\text{in}}e^{-i \widetilde{\omega} r_*}), & r_*\rightarrow -\infty, \\
    (r_\Lambda^{\text{in}} e^{i\omega r_*}, e^{-i\omega r_*}), & r_*\rightarrow +\infty.
    \end{cases}
    \label{eq:Rin}
\end{align}
The factors~$t_\Lambda^{\text{in}}$ and~$r_{\Lambda}^{\text{in}}$ are called the transmission and reflection coefficients, respectively. 
On the other hand, the `up' modes correspond to unit flux of outgoing plane waves at~$\mathscr{H}^-$, with no contribution from~$\mathscr{I}^-$. 
Part of this outgoing flux is transmitted to~$\mathscr{I}^+$ while the other part is reflected back down~$\mathscr{H}^+$, that is,
\begin{align}
    (R_{1,\Lambda}^{\text{up}},R_{2,\Lambda}^{\text{up}}) \sim 
    \begin{cases}
    (e^{i \widetilde{\omega}r_*},r_\Lambda^{\text{up}}e^{-i \widetilde{\omega} r_*}), & r_*\rightarrow -\infty, \\
    (t_\Lambda^{\text{up}} e^{i\omega r_*}, 0), & r_*\rightarrow +\infty.
    \end{cases}
    \label{eq:Rup}
\end{align}

The second basis is the so-called `out-down' basis. 
In particular, the `out' and `down' solutions are the time reversals of the `in' and `out' modes, respectively, so that
\begin{equation}
    R_{1,\Lambda}^{\text{out/down}}=\left( R_{2,\Lambda}^{\text{in/up}} \right)^*, \qquad R_{2,\Lambda}^{\text{out/down}}= \left( R_{1,\Lambda}^{\text{in/up}} \right)^*.
    \label{eq:timereverse}
\end{equation}
In this case, the chosen hypersurface on which we impose the initial behaviour is formed by future null infinity and the future event horizon, $\mathscr{H}^+\cup\mathscr{I}^+$. 
The `out' solutions are outgoing plane waves at~$\mathscr{I}^+$, vanishing at~$\mathscr{H}^+$, such that when they are evolved to the past, part of the wave emanates from~$\mathscr{H}^-$ and part is incoming from~$\mathscr{I}^-$:
\begin{align}
    (R_{1,\Lambda}^{\text{out}},R_{2,\Lambda}^{\text{out}}) \sim 
    \begin{cases}
    (t_\Lambda^{\text{out}} e^{i \widetilde{\omega}r_*},0), &  r_*\rightarrow -\infty, \\
    (e^{i\omega r_*}, r_\Lambda^{\text{out}}e^{-i\omega r_*}), &  r_*\rightarrow +\infty.
    \end{cases}
    \label{eq:Rout}
\end{align}
Analogously, the `down' modes are ingoing solutions near~$\mathscr{H}^+$, vanishing at~$\mathscr{I}^+$, such that in the past, part of the flux is incoming from~$\mathscr{I}^-$ and the other part is outgoing from~$\mathscr{H}^-$, that is,
\begin{align}
    (R_{1,\Lambda}^{\text{down}},R_{2,\Lambda}^{\text{down}}) \sim 
    \begin{cases}
    (r_\Lambda^{\text{down}} e^{i \widetilde{\omega}r_*},e^{-i \widetilde{\omega}r_*}), &  r_*\rightarrow -\infty, \\
    (0, t_\Lambda^{\text{down}}e^{-i\omega r_*}), &  r_*\rightarrow +\infty.
    \end{cases}
    \label{eq:Rdown}
\end{align}
The constants $t_\Lambda^{\text{out/down}}$ and $r_\Lambda^{\text{out/down}}$ are, respectively, the transmission and reflection coefficients for the `out' and `down' modes.

To determine whether classical superradiance exists for charged fermions on RN, it is useful to find relations between the transmission and reflection coefficients of the different modes. We define
\begin{equation}
    W_{1,\Lambda}  =  \widetilde{R}_{1,\Lambda}R_{2,\Lambda} -  \widetilde{R}_{2,\Lambda}R_{1,\Lambda}, \qquad
    W_{2,\Lambda}  =  \widetilde{R}_{1,\Lambda}^*R_{1,\Lambda} -  \widetilde{R}_{2,\Lambda}^*R_{2,\Lambda},
    \label{eq:W1W2}
\end{equation}
for any two pairs of solutions $(R_{1,\Lambda},R_{2,\Lambda})$ and $( \widetilde{R}_{1,\Lambda}, \widetilde{R}_{2,\Lambda})$. It is straightforward to verify that these quantities do not depend on $r$, and as a consequence, 
\begin{align}
    |r_{\Lambda}^{\text{in}}|=|r_{\Lambda}^{\text{up}}|=|r_{\Lambda}^{\text{out}}|=& ~|r_{\Lambda}^{\text{down}}|, \nonumber \\ t_{\Lambda}^{\text{in}}=t_{\Lambda}^{\text{up}}=( t_{\Lambda}^{\text{out}} )^* =&~ ( t_{\Lambda}^{\text{down}} )^*, \nonumber \\
    |r_{\Lambda}^{\text{in}}|^2+|t_{\Lambda}^{\text{in}}|^2=&~1.
    \label{eq:rtrelations}
\end{align}
From~\eqref{eq:rtrelations}, we deduce that all reflection coefficients satisfy the condition~$|r_{\Lambda}|\leq 1$, which confirms the absence of classical superradiance for Dirac fields~\cite{Maeda:1976tm}, similarly to the case of fermions on Kerr black holes~\cite{Chandrasekhar:1985kt}. 
Nonetheless, as we will demonstrate below, due to the frequency shift~\eqref{eq:omegatilde} experienced by an observer near the event horizon, quantum superradiance will occur for fermions in RN backgrounds.

\section{Ambiguities in the notion of quantum vacuum}
\label{sec:quantum}

We now proceed with the canonical quantization of a fermionic field~$\Psi$ on a classical RN background. 
First, we need to choose an orthonormal basis of solutions~$\{\psi_\Lambda\}$ to the Dirac equation~\eqref{eq:Dirac} with respect to the inner product~\eqref{eq:innerproduct}. 
In the last section, we reviewed two well-known possibilities: the `in-up' and the `out-down' bases. 
Second, we need to split our chosen basis into two subsets: one, $\{\psi_\Lambda^+\}$, will generate the one-particle Hilbert space (the space of solutions that we will call particles after quantization), and the other, $\{\psi_\Lambda^-\}$, the one-antiparticle Hilbert space (the modes that we will later call antiparticles). 
Although this procedure is full of ambiguities, it is usual to define the solutions $\psi_\Lambda^+$ and $\psi_\Lambda^-$ as `positive' and `negative' frequency modes, respectively, with respect to a particular time-like coordinate. 
Since RN has a globally time-like Killing vector~$\partial_t$, this provides a natural definition of positive/negative frequency.
However, we will see that even this criterion leaves some freedom in the quantization.

We expand the classical fermionic field~$\Psi$ in terms of the particular chosen basis. 
To define the quantum field operator~$\hat{\Psi}$, we promote the coefficients defining the linear combination to operators on the Fock space:
\begin{equation}
    \hat{\Psi}=\sum_\Lambda \left( \hat{a}_\Lambda \psi_\Lambda^+ + \hat{b}_\Lambda^\dagger \psi_\Lambda^- \right),
    \label{eq:expansionPsi}
\end{equation}
where $\hat{a}_\Lambda$ and $\hat{b}_\Lambda$ are the annihilation operators of the particle and antiparticle sectors, respectively; and their adjoint versions are the associated creation operators. 
We impose the usual anticommutation relations
\begin{equation}
    \{ \hat{a}_\Lambda, \hat{a}_{\Lambda^\prime}^\dagger \} = \{ \hat{b}_\Lambda, \hat{b}_{\Lambda^\prime}^\dagger \} = \delta_{\Lambda,\Lambda'},
    \label{eq:anticommab}
\end{equation}
and all the other anticommutators between creation and annihilation operators vanish. 
The quantum vacuum state~$|0\rangle$ is defined as the state annihilated by all the annihilation operators,
\begin{equation}
    \hat{a}_\Lambda |0\rangle = \hat{b}_\Lambda |0\rangle = 0;
    \label{eq:vacuum}
\end{equation}
that is, the state devoid of particles or antiparticles.
Different choices of basis determine different linear coefficients in the expansion of the Dirac field~$\Psi$, leading to different annihilation and creation operators according to \eqref{eq:expansionPsi}. 
Consequently, this can result in different quantization schemes, each with its own notion of quantum vacuum, particles and antiparticles. 

Our first choice of basis is the `in-up' basis. 
We can define the positive and negative frequency `in' modes with respect to the proper time~$t$ for a static observer asymptotically far from the black hole (at~$\mathscr{I}^-$) with constant azimuthal coordinate~$\varphi$, whose frequency is given by~\mbox{$i\partial_t \psi_\Lambda^{\text{in}}|_\varphi = \omega \psi_\Lambda^{\text{in}}$}. 
We then define~$\psi_\Lambda^{+\text{in}}$ as the modes with positive frequency, $\omega > 0$, and~$\psi_\Lambda^{-\text{in}}$ as those with negative frequency, $\omega < 0$. 
On the other hand, it is also natural to define the frequency of the `up' modes with respect to the sign of the frequency measured by a static observer close to the past event horizon~$\mathscr{H}^-$. 
According to \eqref{eq:R+infty}, the relevant frequency for this observer is not~$\omega$ but the shifted frequency~$ \widetilde{\omega}$~\eqref{eq:omegatilde}. 
As a consequence, we define~$\psi_\Lambda^{+\text{up}}$ as those modes with~$ \widetilde{\omega} > 0$, and~$\psi_\Lambda^{-\text{up}}$ as having $ \widetilde{\omega} < 0$. 
Finally, following the canonical quantization procedure described above, the expansion~\eqref{eq:expansionPsi} translates in this case into
    \begin{multline}
        \hat{\Psi}_{|\text{in}\rangle}=\sum_{l=\frac{1}{2}}^{\infty} \sum_{m=-l}^l \left[ \int_0^{\infty} \text{d}\omega \ \hat{a}^{\text{in}}_\Lambda \psi_\Lambda^{+\text{in}} + \int_{-\infty}^{0} \text{d}\omega \ \hat{b}^{\text{in}\dagger}_\Lambda \psi_\Lambda^{-\text{in}}  
        \right. \\   \left. 
        +\int_0^{\infty} \text{d} \widetilde{\omega} \ \hat{a}^{\text{up}}_\Lambda \psi_\Lambda^{+\text{up}} + \int_{-\infty}^{0} \text{d} \widetilde{\omega} \ \hat{b}^{\text{up}\dagger}_\Lambda \psi_\Lambda^{-\text{up}} \right].
        \label{eq:expansionPsiinup}
    \end{multline}
These annihilation and creation coefficients satisfy the anticommutation relations~\eqref{eq:anticommab}, and define the `in' quantum vacuum, denoted here as~$|\text{in}\rangle$.
This state, by construction, has no incoming flux of particles in the asymptotic past: either coming from~$\mathscr{H}^-$ or~$\mathscr{I}^-$. 

Following similar criteria, we can construct a quantization scheme for the `out-down' basis. 
We choose modes~$\psi_\Lambda^{+\text{out}}$ with positive frequency with respect to a static observer at  future null infinity~$\mathscr{I}^+$ (so that $\omega > 0$), and~$\psi_\Lambda^{+\text{down}}$ to have positive energy with respect to a static observer near the future event horizon~$\mathscr{H}^+$ (giving $ \widetilde{\omega} > 0$). 
Negative frequency modes~$\psi_\Lambda^{-\text{out}}$ and~$\psi_\Lambda^{-\text{down}}$ are defined analogously. 
These choices define the `out' and `down' annihilation and creation operators via the field expansion
    \begin{multline}
        \hat{\Psi}_{|\text{out}\rangle}=\sum_{l=\frac{1}{2}}^{\infty} \sum_{m=-l}^l  \left[ \int_0^{\infty} \text{d}\omega \ \hat{a}^{\text{out}}_\Lambda \psi_\Lambda^{+\text{out}} + \int_{-\infty}^{0} \text{d}\omega \ \hat{b}^{\text{out}\dagger}_\Lambda \psi_\Lambda^{-\text{out}} \right. \\ 
        +  \left. \int_0^{\infty} \text{d} \widetilde{\omega} \ \hat{a}^{\text{down}}_\Lambda \psi_\Lambda^{+\text{down}} + \int_{-\infty}^{0} \text{d} \widetilde{\omega} \ \hat{b}^{\text{down}\dagger}_\Lambda \psi_\Lambda^{-\text{down}} \right],
        \label{eq:expansionPsioutdown}
    \end{multline}
which in turn determine the `out' quantum vacuum~$|\text{out}\rangle$. 
In this vacuum there are no particles in the asymptotic future, either outgoing to~$\mathscr{I}^+$ or going down into~$\mathscr{H}^+$. 

In the case of a charged scalar field, the splitting criterion employed here for both the `in-up' and `out-down' bases is not optional but mandatory~\cite{Balakumar:2022}. 
This requirement arises from the necessity for modes defined to have positive frequency to also have positive Klein-Gordon norm (and negative frequency modes to have negative Klein-Gordon norm) in order to obtain the standard bosonic commutation relations for the creation and annihilation operators. 
Consequently, a `Boulware' state---characterized by the absence of particle flux at both $\mathscr{I}^-$ and $\mathscr{I}^+$---cannot be defined~\cite{Balakumar:2022}. 
In contrast, for fermionic fields, a natural notion of inner product~\eqref{eq:innerproduct} exists without requiring any additional constraints and all modes have positive norm. 
This provides greater flexibility in the choice of mode splitting and, in particular, allows for the definition of a candidate `Boulware' state, as is the case for fermions on rotating black holes~\cite{Casals:2013}.

For instance, we can define the splitting of the `up' modes according to a static observer at~$\mathscr{I}^+$ (instead of with respect to a static observer close to~$\mathscr{H}^-$). 
In this case, the positive frequency modes~$\psi_\Lambda^{+\text{up}}$ would be those with~$\omega >0$ (instead of $ \widetilde{\omega} >0$), while the negative frequency condition would be~$\omega <0$ (and not $ \widetilde{\omega} <0$). By keeping the splitting of the `in' modes with respect to the static observer  at~$\mathscr{I}^-$, this new definition of positive frequency leads to a different quantum vacuum, a `Boulware' state~$|B\rangle$, which has no particles at either~$\mathscr{I}^-$ or~$\mathscr{I}^+$. 
Similarly, we can apply this approach to the `out-down' basis, such that positive frequency `out' and `down' modes are those with~$\omega > 0$, and negative frequency modes are those with~$\omega < 0$. 
This results in another, possibly distinct, `Boulware'-like state~$|B'\rangle$. 

\section{Quantum superradiance} 
\label{sec:quantumsuperradiance}

In this section, we will show that while the `out' state $|\text{out}\rangle$ is empty at future null infinity~$\mathscr{I}^+$ and the future event horizon~$\mathscr{H}^+$, the `in' state $|\text{in}\rangle$ contains an outgoing flux of particles at~$\mathscr{I}^+$ and an ingoing flux transmitted down~$\mathscr{H}^+$. This particle production phenomenon is known as quantum superradiance. 
We will quantify the number of particles per unit time created during this process, as well as the expectation values of the charge current flux~$\hat{J}^r$ and the stress-energy tensor component~$\hat{T}_{r}^{t}$, which corresponds to the energy flux. 
We will see that this particle creation effect results in both the discharge and the energy loss of the black hole.

\subsection{Particle production}

To compare the various notions of quantum vacuum defined in section~\ref{sec:quantum}, we will briefly review the Bogoliubov formalism for fermions in a generic curved spacetime. 

Consider two different orthonormal bases of field modes, $\{\psi_\Lambda^\pm\}$ and~$\{ \widetilde{\psi}_\Lambda^\pm\}$, which define different quantum vacua~$|0 \rangle$ and~$| \widetilde{0} \rangle$, respectively. 
As they are both orthonormal bases of the same Hilbert space of solutions of the Dirac equation, we can write one in terms of the other:
\begin{equation}
     \widetilde{\psi}_\Lambda^+ = \sum_{\Lambda'} \left( \alpha_{\Lambda\Lambda'}^+ \psi_{\Lambda'}^+ + \beta_{\Lambda\Lambda'}^+ \psi_{\Lambda'}^- \right), \qquad \widetilde{\psi}_\Lambda^- = \sum_{\Lambda'} \left( \beta_{\Lambda\Lambda'}^- \psi_{\Lambda'}^+ + \alpha_{\Lambda\Lambda'}^- \psi_{\Lambda'}^- \right).
    \label{eq:basesBogoliubov}
\end{equation}
The constant coefficients from these linear expansions are the Bogoliubov coefficients. 
The orthonormality of the bases and the properties of the Dirac inner product~\eqref{eq:innerproduct} lead us to deduce that the inverse Bogoliubov coefficients are related to those in~\eqref{eq:basesBogoliubov} according to
\begin{equation}
     \widetilde{\alpha}_{\Lambda'\Lambda}^\pm = \left( \alpha_{\Lambda\Lambda'}^\pm \right)^*, \qquad  \widetilde{\beta}_{\Lambda'\Lambda}^\pm = \left( \beta_{\Lambda\Lambda'}^\mp \right)^*.
\end{equation}
Using these relations and taking into account that the expansion of the classical Dirac field~$\Psi$ in terms of both~$\{\psi_\Lambda^\pm\}$ and~$\{ \widetilde{\psi}_\Lambda^\pm\}$ is the same, the relations between the annihilation and creation operators of the two quantization schemes are
\begin{equation}
    \hat{ \widetilde{a}}_\Lambda = \sum_{\Lambda'} \left[ \left( \alpha^+_{\Lambda\Lambda'} \right)^* \hat{a}_{\Lambda'} + \left( \beta_{\Lambda\Lambda'}^+ \right)^* \hat{b}_{\Lambda'}^{\dagger} \right], \qquad
    \hat{ \widetilde{b}}_\Lambda^{\dagger} = \sum_{\Lambda'} \left[ \left( \beta^-_{\Lambda\Lambda'} \right)^* \hat{a}_{\Lambda'} + \left( \alpha_{\Lambda\Lambda'}^- \right)^* \hat{b}_{\Lambda'}^{\dagger} \right].
    \label{eq:abtilde}
\end{equation}

The $\beta$-Bogoliubov coefficients mix the positive and negative frequency modes in the two quantization schemes, and thus, their concepts of particle and antiparticle. 
For example, while~$\hat{ \widetilde{a}}_\Lambda$ annihilates one particle in the `tilde' quantization, in the other this might lead to~$\hat{b}_{\Lambda'}^\dagger$ creating antiparticles. 
Only if all these $\beta$-coefficients vanish, do these notions, as well as the corresponding quantum vacua, coincide. 
This translates to the fact that, relative to the vacuum state~$|0\rangle$, the state~$| \widetilde{0}\rangle$ may contain excitations. 
The total number of these excitations is given by
\begin{equation}
    \mathcal{N} = \sum_\Lambda \langle 0| \hat{ \widetilde{a}}_{\Lambda}^\dagger\hat{ \widetilde{a}}_{\Lambda} + \hat{ \widetilde{b}}_{\Lambda}^\dagger\hat{ \widetilde{b}}_{\Lambda} |0\rangle = \sum_{\Lambda,\Lambda'} \left( |\beta_{\Lambda\Lambda'}^+|^2 + |\beta_{\Lambda\Lambda'}^-|^2 \right),
    \label{eq:numbergeneral}
\end{equation}
where the last equality is a consequence of~\eqref{eq:abtilde} and the anticommutation relations~\eqref{eq:anticommab}. 
We see that coefficients~$\beta_{\Lambda\Lambda'}^+$ and~$\beta_{\Lambda\Lambda'}^-$ provide the number density per mode of fermions and antifermions, respectively.

Returning to the case of a fermionic field on a RN background, the total number of created particles per unit time in the `out' state with respect to the `in' state (and vice versa) is 
\begin{align}
    \mathcal{N}_{|\text{in}\rangle} = \sum_\Lambda \langle \text{in} | &\hat{a}_\Lambda^{\text{out}\dagger} \hat{a}_\Lambda^{\text{out}} + \hat{a}_\Lambda^{\text{down}\dagger} \hat{a}_\Lambda^{\text{down}} 
     +  \hat{b}_\Lambda^{\text{out}\dagger} \hat{b}_\Lambda^{\text{out}} + \hat{b}_\Lambda^{\text{down}\dagger} \hat{b}_\Lambda^{\text{down}}
    | \text{in} \rangle.
\end{align}
Note that, if these two quantum states were the same, we would have $\mathcal{N}_{|\text{in}\rangle}=0$. Let us demonstrate that this is not the case.

From the asymptotic behaviours of the radial functions given in~(\ref{eq:Rin}--\ref{eq:Rdown}), and from the relations between the transmission and reflection coefficients in~\eqref{eq:rtrelations}, we obtain 
\begin{equation}
    R_{j,\Lambda}^{\text{out}} = r_\Lambda^{\text{out}} R_{j,\Lambda}^{\text{in}} + t_\Lambda^{\text{out}} R_{j,\Lambda}^{\text{up}}, \qquad
    R_{j,\Lambda}^{\text{down}} = t_\Lambda^{\text{down}} R_{j,\Lambda}^{\text{in}} + r_\Lambda^{\text{down}} R_{j,\Lambda}^{\text{up}},
\end{equation}
for $j=1,2$. Then, the `in-up' basis is related to the `out-down' basis by the linear combinations
\begin{equation}
    \psi_{\Lambda}^{\text{out}} = r_\Lambda^{\text{out}} \psi_{\Lambda}^{\text{in}} + t_\Lambda^{\text{out}} \psi_{\Lambda}^{\text{up}}, \qquad
    \psi_{\Lambda}^{\text{down}} = t_\Lambda^{\text{down}} \psi_{\Lambda}^{\text{in}} + r_\Lambda^{\text{down}} \psi_{\Lambda}^{\text{up}}.
    \label{eq:relbasisBogoliubov}
\end{equation}
According to \eqref{eq:omegatilde}, when~$qQ>0$, modes with~$\omega>0$ also have ${\widetilde {\omega}}>0$, thus~\eqref{eq:relbasisBogoliubov} is actually a relation between positive frequency solutions for these modes. 
In other words, all the `out' and `down' positive frequency modes are expanded only in terms of positive frequency `in' and `up' modes. 
Consequently, \eqref{eq:basesBogoliubov} implies that the $\beta$-coefficients vanish for~$\omega>0$. For modes with~${\widetilde {\omega}}<0$, which also satisfy~\mbox{$\omega<-qQ/r_+$}, we reach similar conclusions, but with negative frequency solutions. 
However, this does not hold for modes in the range~\mbox{$-qQ/r_+<\omega<0$}, for which ${\widetilde {\omega }}>0$ and we have
\begin{equation}
    \psi_{\Lambda}^{\text{out}-} = r_\Lambda^{\text{out}} \psi_{\Lambda}^{\text{in}-} + t_\Lambda^{\text{out}} \psi_{\Lambda}^{\text{up}+}, \qquad
    \psi_{\Lambda}^{\text{down}+} = t_\Lambda^{\text{down}} \psi_{\Lambda}^{\text{in}-} + r_\Lambda^{\text{down}} \psi_{\Lambda}^{\text{up}+}.
    \label{eq:relbasisBogoliubov+-}
\end{equation}
For this specific frequency interval, there is a mixing of positive and negative frequency modes. 
By comparison with~\eqref{eq:basesBogoliubov}, we have nonvanishing $\beta$-coefficients, which can be identified with the corresponding transmission coefficients. 
Furthermore, due to the equality between the `out' and `down' transmission coefficients~\eqref{eq:rtrelations}, particles and antiparticles are created in equal proportions. 
This is consistent with the fact that particle production is fundamentally a pair creation process, ensuring that the total charge of the produced fermions remains neutral. 
Similarly, for the case where~$qQ<0$, analogous conclusions are drawn, with the only non-vanishing contributions now coming from modes with frequencies within the interval~\mbox{$0< \omega < -qQ/r_+$}. In summary, the total particle production per unit time is
\begin{equation}
    \mathcal{N}_{|\text{in}\rangle} = \frac{1}{16\pi^3} \sum_{l=\frac{1}{2}}^{+\infty} (2l+1) \int_{\min\{-\frac{qQ}{r_+},0\}}^{\max\{-\frac{qQ}{r_+},0\}} \text{d}\omega \ |t_\Lambda^{\text{out}}|^2.
    \label{eq:numberparticles}
\end{equation}
The factor of~$2l+1$ arises from the summation over~$m$, as the transmission coefficients are independent of this quantum number, given that the radial functions are also independent of~$m$~\eqref{eq:R12}. 
Additionally, we have included a prefactor of $1/16\pi^3$, which arises from the normalization of the modes. Although we previously saw that classical superradiance does not occur for fermions on a RN background, this result shows that quantum superradiance is indeed present.
From henceforth, we will call modes with $\omega {\widetilde {\omega }}<0$ `superradiant' modes, since these modes give rise to quantum superradiance.

To numerically compute the transmission coefficients~$t_\Lambda^{\text{out}}$, we solved the system of radial equations~\eqref{eq:R12} with the asymptotic boundary conditions
\begin{equation}
R_{1,\Lambda}^{\text{out}}(r_*) \overset{r_*\to +\infty}{\sim} e^{i\omega r_*}, 
\qquad 
R_{2,\Lambda}^{\text{out}}(r_*) \overset{r_*\to -\infty}{\sim} 0.
\end{equation}
By then calculating the constant~$W_{1,\Lambda}$ in~\eqref{eq:W1W2} for the `out' and `down' radial functions, it follows that 
\begin{equation}
    t_\Lambda^{\text{out}} = \lim_{r_*\rightarrow -\infty} R_{1,\Lambda}^{\text{out}}(r_*) e^{-i \widetilde{\omega}r_*}.
    \label{eq:toutnumerics}
\end{equation}

\begin{figure}
    \centering
    \includegraphics[width=0.7\textwidth]{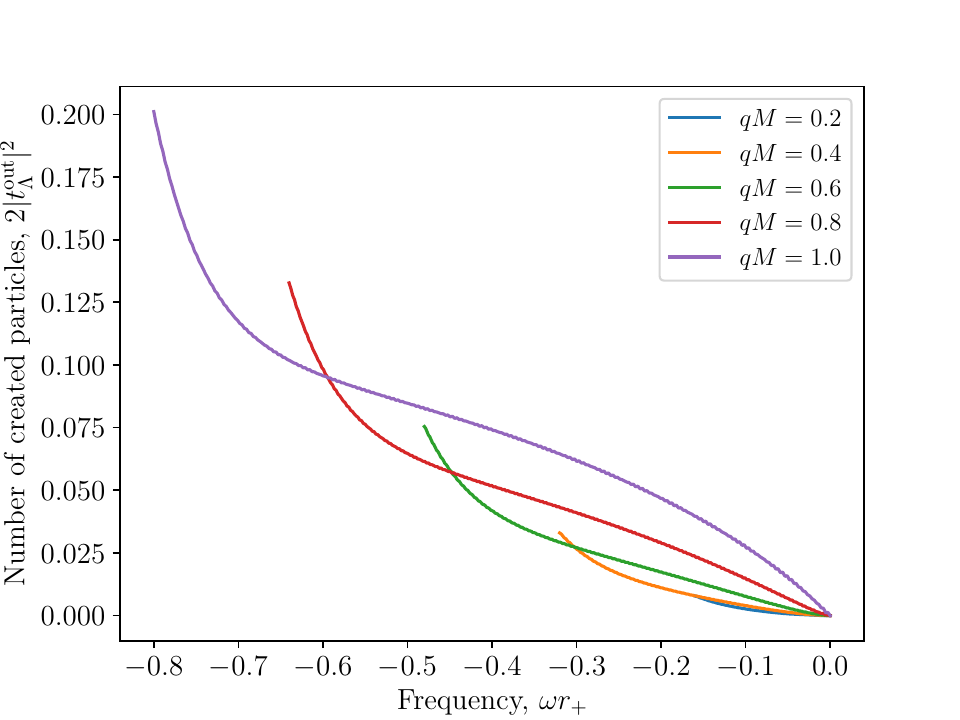}
    \caption{Number of created particles per mode, \mbox{$(2l+1)|t_{\Lambda}^{\text{out}}|^2$}, as a function of the frequency~$\omega$, for modes with $l=1/2$ and various positive fermion charges~$q$. The black hole charge is fixed at $Q=0.8M$. Outside the depicted interval~$[-qQ/r_+,0]$, no particle production occurs, and the particle number drops to zero.}
    \label{fig:densityomega}
\end{figure}

In figure~\ref{fig:densityomega}, we illustrate the dependence of the particle number density on the frequency~$\omega$ for fixed~$l=1/2$, when $qQ>0$. 
Modes with~$\omega = 0$ lack the energy required to cross the transmission barrier, resulting in the reflection of all fermions. 
Consequently, no particle production occurs at vanishing frequencies. 
As the frequency increases in absolute value, the modes gain enough energy to be partially transmitted down the event horizon. 
The particle production peaks at the threshold frequency~$\omega = -qQ/r_+$. 
Beyond this point, the quantum superradiance effect ceases, as described by~\eqref{eq:numberparticles}. 
Our analysis shows that, across all values of~$q$ and~$Q$ studied, the particle density contribution decreases by several orders of magnitude with each increasing value of~$l$. 
For instance, at the threshold~\mbox{$\omega = -qQ/r_+$}, where particle production is most significant,  the contribution of modes with~\mbox{$l>7/2$} is more than ten orders of magnitude smaller than that of the dominant~$l=1/2$ mode. 
Finally, increasing the fermion charge~$q$ broadens the spectrum, leading to an overall enhancement in particle production.
This will be discussed further in section~\ref{sec:energyloss}.

In contrast, consider the `Boulware'-type states~$|B\rangle$ and~$|B'\rangle$, defined in section~\ref{sec:quantum} according to the criterion that all positive frequency modes have~$\omega > 0$ and all negative frequency modes have~$\omega < 0$. 
From~\eqref{eq:relbasisBogoliubov}, it is clear that positive and negative frequencies are not mixed. 
This results in vanishing $\beta$-coefficients across the entire spectrum, meaning that these two states, initially defined using two different bases of field modes, represent the same exact quantization: one that is empty at both~$\mathscr{I}^-$ and~$\mathscr{I}^+$. Therefore, no quantum superradiance occurs with respect to these vacua. 
In the remainder of this paper, we therefore focus on the study of the `in' and `out' quantum vacua.

\subsection{Black hole discharge}

For a Dirac spinor~$\Psi$, we define the classical charge current as
\begin{equation}
J^{\mu} = -q {\overline {\Psi }} \gamma ^{\mu } \Psi.
\label{eq:current}
\end{equation}
This quantity is conserved, $\nabla_\mu J^\mu = 0$. 
To define a charge current quantum operator that takes into account both contributions coming from the flux of particles and antiparticles, we need to introduce a commutator. 
However, in order to preserve the spinorial structure of products of the form~$-q{\overline {\psi }} \gamma ^{\mu } \psi$, this commutator can only act on the annihilation and creation operators and not on the Dirac spinors. 
The quantum charge current operator is thus defined as
\begin{equation}
    \hat{J}^\mu = -\frac{q}{2} \left[ \hat{{\overline {\Psi }}}, \gamma ^{\mu } \hat{\Psi} \right].
\end{equation}
Given a quantization scheme with positive and negative frequency modes~$\psi_\Lambda^+$ and~$\psi_\Lambda^-$, the expectation value of~$\hat{J}^\mu$ is given by
\begin{equation}
    \langle \hat{J}^\mu \rangle = \frac{1}{2}\sum_\Lambda \left( j_\Lambda^{-\mu} - j_\Lambda^{+\mu} \right), \qquad j_\Lambda^{\pm\mu} = -q\psi_\Lambda^\pm\gamma^\mu\psi_\Lambda^\pm.
    \label{eq:expcurrent}
\end{equation}
The expressions for the components of the current~\mbox{$j_\Lambda^\mu = -q\psi_\Lambda\gamma^\mu\psi_\Lambda$} in terms of the functions appearing in the mode ansatz~\eqref{eq:ansatz} are
\begin{align}
    j_\Lambda^t & = \frac{q}{4\pi^2r^2f\sin\theta} \left( |R_{1,\Lambda}|^2S_{1,\Lambda}^2 + |R_{2,\Lambda}|^2S_{2,\Lambda}^2 \right), \nonumber \\
    j_\Lambda^r & = \frac{qL}{4\pi^2r^2\sin\theta} \left( |R_{1,\Lambda}|^2S_{1,\Lambda}^2 - |R_{2,\Lambda}|^2S_{2,\Lambda}^2 \right), \nonumber \\
    j_\Lambda^\theta & = \frac{qL}{2\pi^2r^3\sqrt{f}\sin\theta} \Re \left( R_{1,\Lambda}^* R_{2,\Lambda} \right) S_{1,\Lambda}S_{2,\Lambda}, \nonumber \\
    j_\Lambda^\varphi & = \frac{qL}{2\pi^2r^3\sqrt{f}\sin^2\theta} \Im \left( R_{1,\Lambda}^* R_{2,\Lambda} \right) S_{1,\Lambda}S_{2,\Lambda},
    \label{eq:currentcomp}
\end{align}
where $\Re $ denotes the real part and $\Im $ the imaginary part of complex quantities.

From the properties of the angular functions in Appendix~\ref{sec:propertiesRS}, when performing the finite sum over~$m$ to compute the expectation values~\eqref{eq:expcurrent} we obtain a vanishing contribution from the angular components, so that
\begin{equation}
    \langle \hat{J}^\theta \rangle = \langle \hat{J}^\varphi \rangle = 0,
    \label{eq:angularcurrent}
\end{equation}
independently of the quantum state under consideration. 
This is expected from the spherical symmetry of the configuration. 

Since we want to quantify quantum superradiance, we are particularly interested in the components of the charge current leading to different expectation values for `in' and `out' states. 
Using that the `out-down' basis is the time reverse of the `in-out' basis (see~\eqref{eq:timereverse}) and the symmetries of the angular functions under the transformation~$m\rightarrow -m$ given in Appendix~\ref{sec:propertiesRS} (the radial functions are independent of~$m$), we arrive at the results
\begin{equation}
    \langle \text{in} | \hat{J}^t | \text{in} \rangle = \langle \text{out} | \hat{J}^t | \text{out} \rangle, \qquad
    \langle \text{in} | \hat{J}^r | \text{in} \rangle = - \langle \text{out} | \hat{J}^r | \text{out} \rangle.
    \label{eq:currentB-B+}
\end{equation}
Thus, we will now focus on the computation of the expectation value of the radial component of the charge current. 

From the semiclassical Maxwell equation \mbox{$\nabla_\mu F^{\mu\nu} = \langle \hat{J}^\nu \rangle$}, where \mbox{$F^{\mu\nu}=\partial^\mu A^\nu - \partial^\nu A^\mu$} is the antisymmetric electromagnetic tensor, we deduce that the expectation value of the current density operator is conserved for all quantum vacua: $\nabla_\mu \langle \hat{J}^\mu \rangle = 0$. Taking into account the vanishing angular components~\eqref{eq:angularcurrent} and the fact that the quantum states considered here are static, integrating this conservation equation leads to
\begin{equation}
    \langle \hat{J}^r \rangle = -\frac{\mathcal{K}}{r^2},
\end{equation}
where~$\mathcal{K}$ is an integration constant independent of $r$. 
From~\eqref{eq:expcurrent}, the sign of~$\mathcal{K}$ matches that of the contribution to the charge current from the positive frequency modes, $j_\Lambda^{+\mu}$. Consequently, $\mathcal{K}$ represents the net charge flux emitted by the black hole, defined as the charge flux of particles minus that of antiparticles. 

To compute~$\mathcal{K}$ for the `in' vacuum we only need to evaluate~$\langle \hat{J}^r \rangle$ at~$r_*\rightarrow +\infty$. 
Using the asymptotics of the radial functions~(\ref{eq:Rin}--\ref{eq:Rup}), this results in
\begin{equation}
    \mathcal{K}_{|\text{in}\rangle} = \frac{q}{16\pi^3} \sum_{l=\frac{1}{2}}^{+\infty} (2l+1) \int_{-\frac{qQ}{r_+}}^0 \text{d}\omega \ |t_\Lambda^{\text{out}}|^2.
    \label{eq:K}
\end{equation}
Note that when $qQ<0$, the lower limit of the integral is larger than the upper limit, introducing a negative sign when the order is reversed. 
Only the superradiant modes contribute to the charge flux, with the absolute value of $\mathcal{K}_{|\text{in}\rangle}$ given by the total particle number per unit time in~\eqref{eq:numberparticles} multiplied by the charge of the fermionic field~$q$. 
This result is independent of the chirality~$L$ due to the invariance of $\sum_\Lambda j_\Lambda^r$~\eqref{eq:currentcomp} under the transformation \mbox{$R_{1,\Lambda} \leftrightarrow R_{2,\Lambda}$}.

The contribution of particles to the total charge current is equal in magnitude but opposite in sign to that of antiparticles;  in other words \mbox{$\sum_\Lambda j_\Lambda^{+r} = - \sum j_\Lambda^{-r}$}. 
In addition, from~\eqref{eq:K}, we observe that the sign of the charge flux~$\mathcal{K}_{|\text{in}\rangle}$ matches the sign of the black hole charge~$Q$. 
This implies that when the black hole is positively (negatively) charged, positive (negative) charges are emitted outwards while an equal number of negative (positive) charges are absorbed, resulting in the discharge of the black hole due to quantum superradiance.

Finally, the expression for~$\mathcal{K}$ in~\eqref{eq:K} is only valid for the `in' state (for `out' state, according to~\eqref{eq:currentB-B+}, we have \mbox{$\mathcal{K}_{|\text{out}\rangle}=-\mathcal{K}_{|\text{in}\rangle}$}), and each quantum vacuum has its own value of~$\mathcal{K}$. 
For the `Boulware'-type quantum state~$|B\rangle$, we find that $\mathcal{K}_{|B\rangle}=0$ and there is no charge current in the radial direction.

\subsection{Black hole energy loss}
\label{sec:energyloss}

The classical stress-energy momentum tensor for a Dirac field~$\Psi$ in background gravitational and electromagnetic fields is
\begin{equation}
T_{\mu \nu }= \frac {i}{2} \left[ {\overline {\Psi }} \gamma _{(\mu }
\nabla _{\nu )} \Psi - \left( \nabla _{( \mu }{\overline {\Psi }}\right)
\gamma _{\nu )} \Psi + 2iqA_{(\mu} {\overline {\Psi }} \gamma_{\nu) } \Psi \right],
\label{eq:Tmunuclassical}
\end{equation}
where parentheses are used to denote symmetrization of indices. 
With the same caveat as for the current, namely that commutators act only on the annihilation and creation operators and not the Dirac spinors, the associated quantum operator is
\begin{equation}
    \hat{T}_{\mu\nu} = \frac{i}{4}  \left\{ \left[ \hat{\overline{\Psi}}, \gamma_{(\mu} \nabla_{\nu)} \hat{\Psi} \right] - \left[ \nabla_{(\mu} \hat{\overline{\Psi}}, \gamma_{\nu)} \hat{\Psi} \right]  + 2iqA_{(\mu} \left[ \hat{{\overline {\Psi }}}, \gamma_{\nu)} \hat{\Psi} \right] \right\}.
\end{equation}
Fixing a particular quantization scheme having positive and negative frequency modes~$\psi_\Lambda^+$ and~$\psi_\Lambda^-$, we find the expectation value of $\hat{T}_{\mu\nu}$ to be
\begin{equation}
    \langle \hat{T}_{\mu\nu} \rangle = \frac{1}{2} \sum_\Lambda \left( t_{\mu\nu,\Lambda}^- - t_{\mu\nu,\Lambda}^+ \right),
\end{equation}
where $t_{\mu\nu,\Lambda}^\pm$ are the classical stress-energy momentum tensor components~\eqref{eq:Tmunuclassical} for the modes~$\psi_\Lambda^\pm$. 
Their expressions (omitting~$\Lambda$ and the~$\pm$ signs for simplicity in the notation) in terms of the radial and angular functions  and mode contributions to the charge current~\eqref{eq:currentcomp} are
 \begin{align}
    t_{tt} =& ~ -\left( \omega+\frac{qQ}{r} \right) \frac{j_t}{q}, \nonumber \\
    t_{tr} =& ~ -\frac{1}{2} \left( \omega+\frac{qQ}{r} \right)\frac{j_r}{q} - \frac{1}{8\pi^2r^2\sin\theta}\left[ \Im\left( R_1^*R_1^\prime \right) S_1^2 + \Im\left( R_2^*R_2^\prime \right) S_2^2 \right], \nonumber \\
    t_{t\theta} =& ~ -\frac{1}{2} \left( \omega+\frac{qQ}{r} \right)\frac{j_\theta}{q} - \frac{L}{4r\sin\theta} \left( \frac{r}{2}f^\prime - f \right) \frac{j_\varphi}{q}, \nonumber \\
    t_{t\varphi} =& ~ \frac{m}{2}\frac{j_t}{q} + \frac{L\cos\theta f}{4} \frac{j_r}{q} + \frac{L\sin\theta}{4r} \left( \frac{r}{2}f^\prime - f \right) \frac{j_\theta}{q} - \frac{1}{2}\left( \omega+\frac{qQ}{r} \right)\frac{j_\varphi}{q}, \nonumber \\
    t_{rr} =& ~ \frac{L}{4\pi^2r^2f\sin\theta} \left[ \Im\left( R_1^* R_1^\prime \right) S_1^2 - \Im\left( R_2^* R_2^\prime \right) S_2^2 \right], \nonumber \\
    t_{r\theta} =& ~ \frac{L}{8\pi^2r\sqrt{f}\sin\theta} \left[ \Im\left( R_1^* R_2^\prime \right) + \Im\left( R_2^* R_1^\prime \right)\right] S_1 S_2, \nonumber \\
    t_{r\varphi} =& ~ -\frac{L\cos\theta}{4 f} \frac{j_t}{q} + \frac{m}{2} \frac{j_r}{q} - \frac{L}{8\pi^2r\sqrt{f}} \left[ \Re\left( R_1^* R_2^\prime \right) - \Re\left( R_2^* R_1^\prime \right) \right] S_1 S_2, \nonumber \\
    t_{\theta\theta} =& ~ -\frac{L}{4\pi^2 r\sqrt{f}\sin\theta} \Im\left( R_1^*R_2 \right) \left( S_1^\prime S_2 - S_1 S_2^\prime \right), \nonumber \\
    t_{\theta\varphi} =& ~ \frac{m}{2} \frac{j_\theta}{q} + \frac{L}{8\pi^2r\sqrt{f}} \Re\left( R_1^* R_2 \right) \left( S_1^\prime S_2 - S_1 S_2^\prime \right), \nonumber \\
    t_{\varphi\varphi} =& ~ m \frac{j_\varphi}{q}.
    \label{eq:tmunu}
\end{align}

Taking into account the properties of the angular functions in Appendix~\ref{sec:propertiesRS}, substituting the modes \eqref{eq:ansatz} in the stress-energy momentum tensor components in~\eqref{eq:tmunu}, and performing the finite sum over $m$, all components of the stress-energy momentum tensor expectation value vanish except for~$\langle \hat{T}_{tt} \rangle$, $\langle \hat{T}_{rr} \rangle$ and~$\langle \hat{T}_{tr} \rangle$. 
However, while the first two coincide for the `in' and `out' states, this is not the case for the $tr$-component, which satisfies
\begin{equation}
    \langle \text{in} | \hat{T}_{tr} | \text{in} \rangle = - \langle \text{out} | \hat{T}_{tr} | \text{out} \rangle.
\end{equation}
In order to quantify the quantum superradiance phenomenon we focus now on calculating this radial energy flux expectation value.

Due to the electromagnetic background, the expectation value of the stress-energy momentum tensor is not conserved: $\nabla^\mu \langle \hat{T}_{\mu\nu} \rangle = 4\pi F_{\mu\nu} \langle \hat{J}^{\mu} \rangle$. Taking into account the fact that all the expectation values are time-independent, we integrate the equation for~$\nu=t$, resulting in 
\begin{equation}
    \langle \hat{T}_t^r \rangle = -\frac{\mathcal{L}}{r^2} + \frac{4\pi \mathcal{K}Q}{r^3},
    \label{eq:TtrL}
\end{equation}
where~$\mathcal{L}$ does not depend on $r$, but does depend on the particular quantum state considered. 
Physically ${\mathcal{L}}$ is the flux of energy from the black hole. 
To evaluate~$\langle \hat{T}_t^r \rangle$ for the `in' vacuum at~$r_* \rightarrow +\infty$, we use the asymptotic behaviour of the radial functions~(\ref{eq:Rin}--\ref{eq:Rup}), as well as the properties of the angular functions given in Appendix~\ref{sec:propertiesRS}. 
Identifying this result with~\eqref{eq:TtrL}, we obtain 
\begin{equation}
    \mathcal{L}_{|\text{in}\rangle} = -\frac{1}{16\pi^3} \sum_{l=\frac{1}{2}}^{+\infty} (2l+1) \int_{-\frac{qQ}{r_+}}^0 \text{d}\omega \ \omega |t_\Lambda^{\text{out}}|^2.
\end{equation}
When $qQ<0$, we again need to reverse the order of the integral limits and introduce a negative sign. Each superradiant mode contributes to the energy flux in the radial direction with an energy proportional to its particle number, $(2l+1)|t_\Lambda^{\text{out}}|^2$, and its frequency~$\omega$. 
Due to the invariance of $\sum_\Lambda t_{tr,\Lambda}$ under the exchange of $R_{1,\Lambda}$ and $R_{2,\Lambda}$~\eqref{eq:tmunu}, the energy flux~$\mathcal{L}_{|\text{in}\rangle}$ is the same for positive and negative chiralities. 

\begin{figure}
    \centering
    \includegraphics[width=0.7\textwidth]{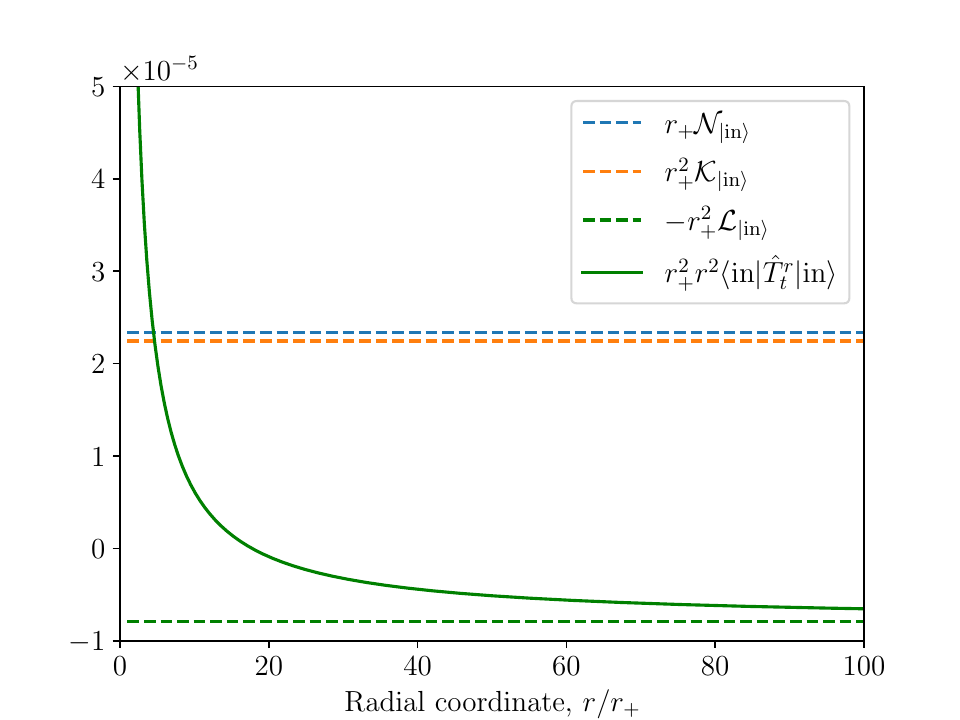}
    \caption{Expected energy density flux dissipated by the black hole, $r^2 \langle \text{in} | \hat{T}_t^r | \text{in} \rangle$, as a function of the radial coordinate, for $q M = 0.6$ and $Q = 0.8 M$. 
    Far from the black hole, it approaches a horizontal asymptote at~$-\mathcal{L}_{|\text{in}\rangle}$. 
    The total number of particles created per time, $\mathcal{N}_{|\text{in}\rangle}$, and the charge flux constant, $\mathcal{K}_{|\text{in}\rangle}$, are also shown.}
    \label{fig:NKLr}
\end{figure}

In particular, $\mathcal{L}_{|\text{in}\rangle}$ is always positive, and due to the black hole discharge studied above, we have $\mathcal{K}_{|\text{in}\rangle}Q > 0$. 
As a result, there is a spherical surface with radius
\begin{equation}
    r_0=\frac{4\pi \mathcal{K}_{|\text{in}\rangle} Q}{\mathcal{L}_{|\text{in}\rangle}},
    \label{eq:r0}
\end{equation}
where the expectation value in~\eqref{eq:TtrL} vanishes. Inside this sphere, there is an ingoing flux of energy into the black hole, while outside, there is a net energy loss. This can be seen in figure~\ref{fig:NKLr}, which shows how the energy flux~$r^2\langle \text{in} | \hat{T}_t^r | \text{in} \rangle$ decreases as one moves away from the black hole, and asymptotically approaches the constant $-\mathcal{L}_{|\text{in}\rangle}$. We also show the total particle number per unit time $\mathcal{N}_{|\text{in}\rangle}$~\eqref{eq:numberparticles}, and the charge flux constant $\mathcal{K}_{|\text{in}\rangle}$~\eqref{eq:K}. The behaviour of~$r^2\langle \text{in} | \hat{T}_t^r | \text{in} \rangle$ resembles that observed in the case of a charged scalar field on RN~\cite{Balakumar:2020gli}, where an `effective ergosphere' indicates a sign change in this component of the stress-energy momentum tensor outside the event horizon~\cite{DiMenza:2014vpa,Denardo:1973pyo,Denardo:1974qis}. 

\begin{figure}
    \centering
    \includegraphics[width=0.7\textwidth]{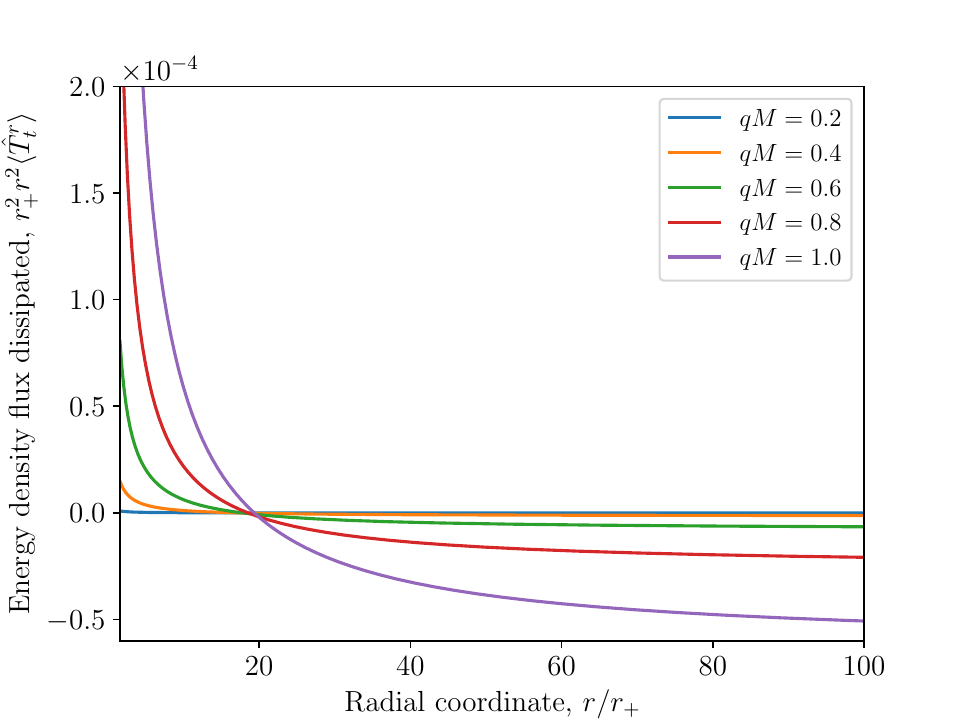}
    \caption{Expected energy density flux dissipated by the black hole, $r^2 \langle \text{in} | \hat{T}_t^r | \text{in} \rangle$, as a function of the radial coordinate, for $Q = 0.8 M$ and various fermion charges $q$.}
    \label{fig:Ttr}
\end{figure}

\begin{figure}
    \centering
    \includegraphics[width=0.7\textwidth]{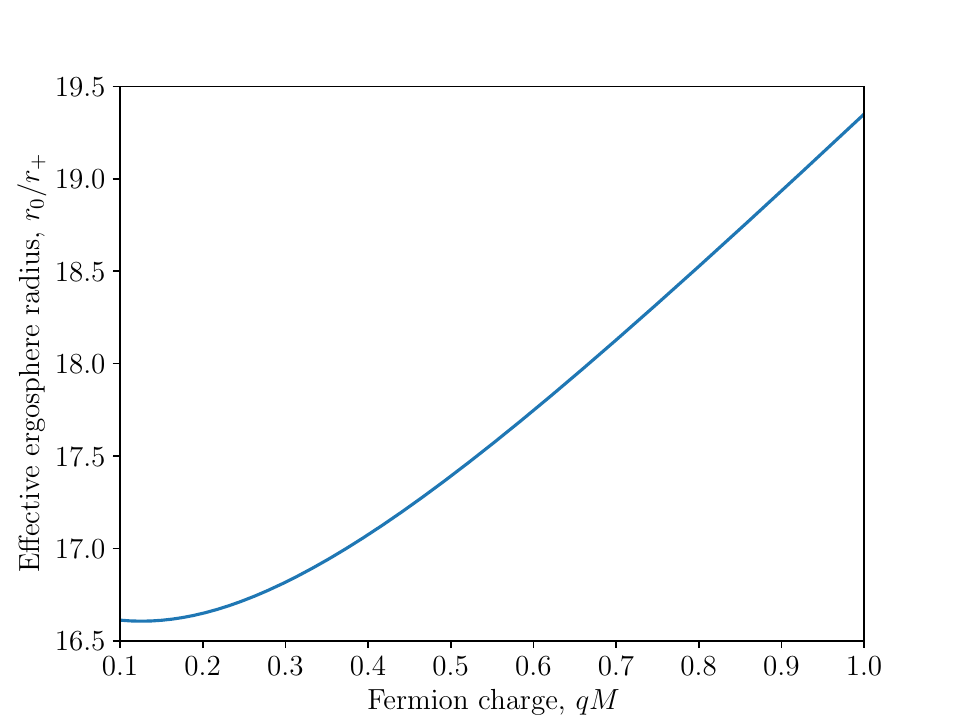}
    \caption{Effective ergosphere radius $r_{0}$, where the expectation value $r^2 \langle \text{in} | \hat{T}_t^r | \text{in} \rangle$ vanishes, as a function of the fermion field charge $q$, with $Q = 0.8M$.}
    \label{fig:ergo}
\end{figure}

\begin{figure}
    \centering
    \includegraphics[width=0.7\textwidth]{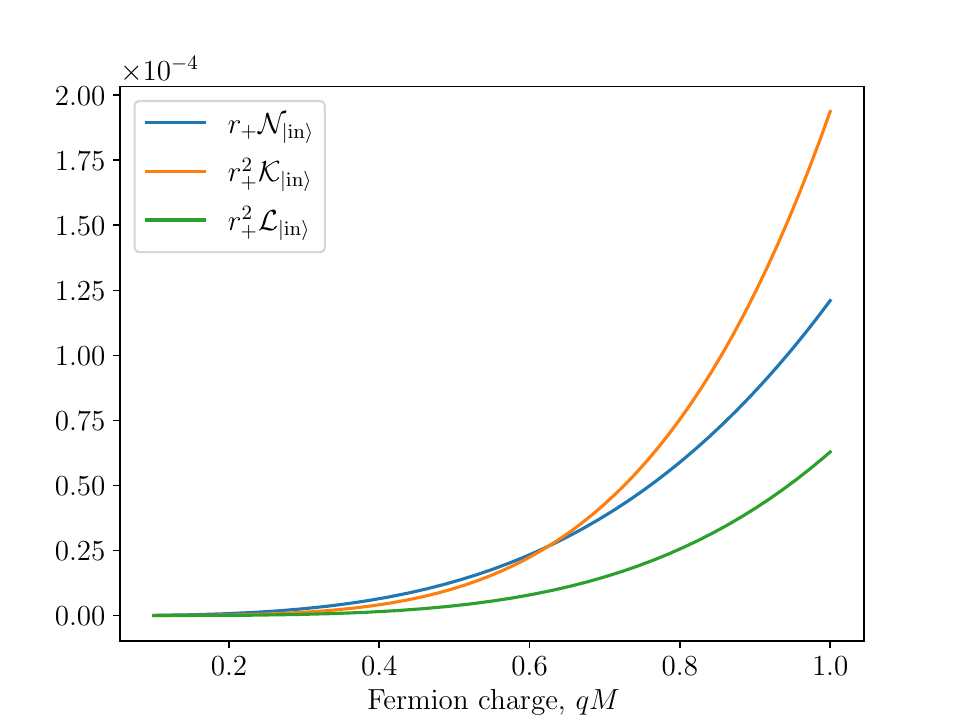}
    \caption{Enhancement of quantum superradiance when increasing the fermion charge~$q$, for a black hole charge \mbox{$Q = 0.8 M$}. We show the quantities $\mathcal{N}_{|\text{in}\rangle}$, $\mathcal{K}_{|\text{in}\rangle}$ and~$\mathcal{L}_{|\text{in}\rangle}$ as functions of $q$.}
    \label{fig:NKLq}
\end{figure}

In figure~\ref{fig:Ttr} we explore how~$r^2\langle \text{in} | \hat{T}_t^r | \text{in} \rangle$ changes as we vary the fermion charge~$q$. As the charge~$q$ increases, the ingoing energy flux inside the effective ergosphere grows, enabling a greater extraction of energy from the black hole, which is expelled outside the ergosphere. This results in a net energy gain at the expense of drawing energy from the black hole. Figure~\ref{fig:ergo} shows how the boundary $r_0$ \eqref{eq:r0} of the effective ergosphere shifts with $q$, revealing a slight expansion of the ergosphere as $q$ increases. This expansion enhances quantum charge superradiance. Indeed, in figure ~\ref{fig:NKLq} we observe that particle creation~$\mathcal{N}_{|\text{in}\rangle}$, charge flux~$\mathcal{K}_{|\text{in}\rangle}$ and energy flux~$\mathcal{L}_{|\text{in}\rangle}$ all increase with larger~$q$, as the electromagnetic interaction between the RN black hole and the charged field intensifies.

Although these results are similar to the scalar case, charge quantum superradiance is considerably more intense for fermions. Notably, the effective ergosphere is one order of magnitude larger for fermions than for charged scalars, for which $r_{0}\sim 2r_{+}$~\cite{Balakumar:2020gli}. This leads to charge and energy fluxes that are two orders of magnitude higher than in the scalar case.

We close our analysis of quantum superradiance for charged fermions by considering how the quantities $\mathcal{N}_{|\text{in}\rangle}$, $\mathcal{K}_{|\text{in}\rangle}$ and~$\mathcal{L}_{|\text{in}\rangle}$
depend on the signs of the fermion charge $q$ and the black hole charge $Q$. In figure~\ref{fig:NKLq} we consider $q>0$ and $Q>0$.
Under the transformation~\mbox{$qQ \rightarrow -qQ$}, both the total number of particles (per unit time)~$\mathcal{N}_{|\text{in}\rangle}$ and the energy flux~$\mathcal{L}_{|\text{in}\rangle}$ remain invariant. In contrast, the charge flux~$\mathcal{K}_{|\text{in}\rangle}$ changes sign when~\mbox{$Q\rightarrow -Q$} but remains invariant when~\mbox{$q\rightarrow -q$}. 
This behaviour follows from the transformation of the radial differential equation~\eqref{eq:R12}  under the mapping~\mbox{$qQ \rightarrow -qQ$}, leading to the transformation~\mbox{$R_{j,(\omega,l,m)}^{\text{out}} \rightarrow (R_{j,(-\omega,l,m)}^{\text{out}})^*$}. 
Consequently, as implied by~\eqref{eq:toutnumerics}, the transmission coefficient transforms as~\mbox{$t_{(\omega,l,m)}^{\text{out}} \rightarrow (t_{(-\omega,l,m)}^{\text{out}})^*$}.
Therefore the behaviour for different signs of $q$ and $Q$ can be deduced from that depicted in figure~\ref{fig:NKLq} by making an appropriate transformation.

\section{Conclusions}
\label{sec:conc}

In this paper we have studied a massless, charged fermion field propagating on a static, charged RN black hole background.
Classically, the charged fermion field does not exhibit superradiance \cite{Maeda:1976tm}, but we have shown that the quantum analogue of charge superradiance can occur, depending on the quantum state of the field.  
We define an `in' vacuum state which is empty at both the past horizon and past null infinity. 
In this state, quantum superradiance is present: charged fermions are spontaneously emitted into those field modes whose frequency lies in the range for which a bosonic field would exhibit classical superradiance.
As a result, the black hole discharges and also loses energy. 

However, there is an inherent ambiguity in how the vacuum state is defined; the `in' vacuum is not the only possibility.
For example, we can construct the time-reverse of the `in' state, namely the `out' vacuum,  which is as empty as possible at both the future event horizon and future null infinity.
Both these states can be defined analogously for a quantum charged scalar field on RN \cite{Balakumar:2020gli}.
For a quantum charged fermion as considered here, there is a third possibility. 
We can define a state which is as empty as possible at {\em {both}} past and future null infinity.
Such as state can only be defined for fermions; there is no analogue for a charged scalar field \cite{Balakumar:2022}.
In this `Boulware'-like state, there is no spontaneous emission of charged fermions, and accordingly this state is that which most closely resembles the `Boulware' state  \cite{Boulware:1975} for a neutral scalar or fermion field on a static Schwarzschild or RN black hole. 

The situation for charge superradiance on charged black holes, as studied in this paper, is somewhat analogous to that for rotational superradiance on rotating Kerr black holes.
In both scenarios, classical superradiance is present for scalars but not for fermions; however both scalars and fermions can exhibit quantum superradiance.
Furthermore, for scalar fields in both set-ups, it is not possible to define a `Boulware'-like state which is as empty as possible at both past and future null infinity \cite{Ottewill:2000,Balakumar:2022}.
Considering neutral fermions on a rotating Kerr black hole, as is the case here for charged fermions on a charged black hole, a `Boulware'-like state can be defined \cite{Casals:2013}. 
However, while this state on Kerr is a vacuum state asymptotically far from the black hole, it diverges on the stationary limit surface (the boundary of the ergosphere) \cite{Casals:2013}. 
It would be interesting to investigate whether the `Boulware'-like state we have defined here in this paper is regular everywhere outside the event horizon. 
This would require a study of all the components of the charge current and stress-energy momentum tensor, which is beyond the scope of our present work.

\appendix

\section{Dirac formalism on RN background}
\label{sec:useful}

The Dirac matrices $\gamma ^{\mu }$ satisfy the anticommutation relations~$\{\gamma^\mu,\gamma^\nu\} = 2g^{\mu \nu }$, where $g^{\mu \nu }$ is the inverse metric.
A suitable basis of $\gamma ^{\mu }$ matrices for the RN metric (\ref{eq:RNmetric}) is given by \cite{Unruh:1974bw}
\begin{equation}
\gamma ^{t}   = 
\frac {1}{{{\sqrt {f(r)}}}} { \widetilde {\gamma }}^{0},
\qquad 
\gamma ^{r}  = 
{\sqrt {f(r)}} { \widetilde {\gamma }}^{3}, \qquad
\gamma ^{\theta } = 
\frac {1}{r} { \widetilde {\gamma }^{1}},
\qquad
\gamma ^{\varphi }  = 
\frac {1}{r\sin \theta } { \widetilde {\gamma }}^{2},
\label{eq:gamma}
\end{equation}
where the selected representation for the flat-space matrices~${ \widetilde {\gamma }}^{a}$ is
\begin{equation}
{ \widetilde {\gamma }}^{0} =
\left(
\begin{array}{cc}
iI_{2} & 0  \\
0 & -iI_{2}
\end{array}
\right) ,
\qquad
{ \widetilde {\gamma }}^{j} =
\left(
\begin{array}{cc}
0 & i\sigma _{j} \\
-i\sigma _{j} & 0
\end{array}
\right) ,
\label{eq:flatspacegamma}
\end{equation}
with $I_{2}$ the $2\times 2$ identity matrix and $\sigma _{i}$  the usual Pauli matrices
\begin{equation}
\sigma _{1} =
\left(
\begin{array}{cc}
0 & 1 \\
1 & 0
\end{array}
\right) ,
\quad
\sigma _{2} =
\left(
\begin{array}{cc}
0 & -i \\
i & 0
\end{array}
\right) ,
\quad
\sigma _{3}=
\left(
\begin{array}{cc}
1 & 0 \\
0 & -1
\end{array}
\right) .
\end{equation}
The chirality matrix in~\eqref{eq:lefthanded} is
\begin{equation}
\gamma ^{5} =
i { \widetilde {\gamma }}^{0} { \widetilde {\gamma }}^{1} { \widetilde {\gamma }}^{2}
{ \widetilde {\gamma }}^{3} =
\left(
\begin{array}{cc}
0 & I_{2}  \\
I_{2} & 0
\end{array}
\right).
\label{eq:gamma5}
\end{equation}
In the above, Latin indices~$a$ represent Minkowski coordinates, while Greek indices~$\mu$ denote the coordinates used in~\eqref{eq:RNmetric} to describe the RN black hole. 

The spinor covariant derivatives~$\nabla _{\mu }$ in the Dirac equation~\eqref{eq:Dirac} are defined according to~\cite{Unruh:1974bw}
\begin{equation}
\nabla _{\mu } \Psi = \frac {\partial }{\partial x^{\mu }} \Psi - \Gamma _{\mu }\Psi ,
\end{equation}
where~$\Gamma_{\mu }$ are the spinor connection matrices, given in the RN background by
\begin{eqnarray}
\Gamma _{t} & = &
\frac {1}{4} \frac{\text{d} f}{\text{d} r}
{ \widetilde {\gamma }}^{0} { \widetilde {\gamma }}^{3},
\nonumber \\
\Gamma _{r}& =  & 0,
\nonumber \\
\Gamma _{\theta} & = &
-\frac{1}{2} \sqrt{f(r)} { \widetilde {\gamma }}^{1}
{ \widetilde {\gamma }}^{3}
,
\nonumber \\
\Gamma _{\phi}& = &
-\frac{1}{2} \left[ \sqrt{f(r)} \sin\theta \, { \widetilde {\gamma }}^{2} { \widetilde {\gamma }}^{3} + \cos\theta \,
{ \widetilde {\gamma }}^{2} { \widetilde {\gamma }}^{1} \right].
 \label{eq:Gamma}
\end{eqnarray}
This leads to the expression for the Dirac equation in~\eqref{eq:DiracCoord}. 
The spinor covariant derivative of the conjugate spinor, defined as~${\overline {\Psi }} = \Psi ^{\dagger } \widetilde{\gamma}^0$, with~$\Psi ^{\dagger }$ being the hermitian conjugate of~$\Psi $, is 
\begin{equation}
    \nabla_{\mu}\overline{\Psi} = \partial_\mu \overline{\Psi}+\overline{\Psi}\Gamma_\mu .
\end{equation}

\section{Properties of the angular functions}
\label{sec:propertiesRS}

The angular functions~$S_{1,\Lambda}(\theta)$ and~$S_{2,\Lambda}(\theta)$ are related to the well-known spin-weighted spherical harmonics $_sY_l^m(\theta,\varphi)$ \cite{Newman:1966ub,Goldberg:1966uu}:
\begin{align}
    S_{1,\Lambda}(\theta) & = \sqrt{\sin\theta} _{\frac{1}{2}}Y_l^{-m}(\theta,\varphi) e^{im\varphi}, \nonumber \\ S_{2,\Lambda}(\theta) & = \sqrt{\sin\theta} _{-\frac{1}{2}}Y_l^{-m}(\theta,\varphi) e^{im\varphi}.
\end{align}
They  are normalized according to
\begin{equation}
    \int_0^\pi S_{1,\Lambda}(\theta) \ \text{d}\theta = \int_0^\pi S_{2,\Lambda}(\theta) \ \text{d}\theta = 1
\end{equation}
and satisfy the following addition relations, which can easily be deduced from those for the spin-weighted spherical harmonics~\cite{Monteverdi:2024xyp}:
\begin{align}
    \sum_{m=-l}^l S_{1,\Lambda}(\theta)^2 =& \sum_{m=-l}^l S_{2,\Lambda}(\theta)^2  = \frac{2l+1}{4\pi} \sin\theta, \nonumber \\
    \sum_{m=-l}^l m S_{1,\Lambda}(\theta)^2 =& -\sum_{m=-l}^l m S_{2,\Lambda}(\theta)^2  = \frac{2l+1}{8\pi} \sin\theta \cos\theta, \nonumber \\
    \sum_{m=-l}^l S_{1,\Lambda}(\theta) S_{2,\Lambda}(\theta) =& \sum_{m=-l}^l m S_{1,\Lambda}(\theta) S_{2,\Lambda}(\theta)  = 0, \nonumber \\
    \sum_{m=-l}^l S_{j,\Lambda}(\theta)\frac{\text{d}}{\text{d}\theta} S_{k,\Lambda}(\theta) =& ~ 0,
    \label{eq:angularproperties}
\end{align}
for $j,k=1,2$. Another useful property concerns the symmetries of the angular functions under the transformation~\mbox{$m \rightarrow -m$}. 
From their governing equations~\eqref{eq:S12} we have
\begin{equation}
    S_{1,(-m,l,\omega)}=\pm S_{2,(m,l,\omega)}, \qquad S_{2,(-m,l,\omega)}=\mp S_{1,(m,l,\omega)}.
    \label{eq:symmetriesm}
\end{equation}

\acknowledgments

The authors would like to thank
Mercedes Mart\'{\i}n-Benito and Luis J.~Garay for helpful discussions. A.A.D.~acknowledges support through the Grant PID2020-118159GB-C44 and PID2023-149018NB-C44 (funded by MCIN/AEI/10.13039/501100011033). The work of E.W.~is supported by STFC grant number ST/X000621/1.
Data supporting this publication can be freely downloaded from the University of Sheffield research data repository at {\url 
{https://doi.org/10.15131/shef.data.27350052}}, under the terms of the Creative Commons Attribution (CC-BY) licence.

\bibliographystyle{JHEP}
\bibliography{fermion}

\end{document}